\documentclass[conference]{IEEEtran}
\usepackage{todonotes}
\usepackage{graphicx}
\usepackage{subfigure}
\usepackage{amsmath}
\usepackage{amsthm}
\usepackage{url}
\usepackage{xcolor,colortbl}
\usepackage{tabularx}
\usepackage{multirow}
\usepackage{enumitem}
\newlist{Properties}{enumerate}{2}
\setlist[Properties]{label=Property \arabic*.,itemindent=*}

\IEEEoverridecommandlockouts
\begin{document}

\title{Autonomous Vehicle: Security by Design}

\author{\IEEEauthorblockN{Anupam Chattopadhyay and Kwok-Yan Lam$^\dagger$}
\IEEEauthorblockA{School of Computer Science and Engineering,
Nanyang Technological University, Singapore\\
\{anupam, kwokyan.lam\}@ntu.edu.sg}\thanks{$^\dagger$Corresponding author.}}

\maketitle

\begin{abstract}
Security of (semi)-autonomous vehicles is a growing concern, first, due to the increased exposure of the functionality to the potential attackers; second, due to the reliance of car functionalities on diverse (semi)-autonomous systems; third, due to the interaction of a single vehicle with myriads of other smart systems in an urban traffic infrastructure. Beyond these technical issues, we argue that the security-by-design principle for smart and complex autonomous systems, such as an Autonomous Vehicle (AV) is poorly understood and rarely practiced. Unlike traditional IT systems, where the risk mitigation techniques and adversarial models are well studied and developed with security design principles such as security perimeter and defence-in-depth, the lack of such a framework for connected autonomous systems is plagueing the design and implementation of a secure AV. We attempt to identify the core issues of securing an AV. This is done methodically by developing a security-by-design framework for AV from the first principle. Subsequently, the technical challenges for AV security are identified.
\end{abstract}

\section{Introduction}
Modern cars are fitted with a range of autonomy features. In order to distinguish such cars with varying degrees of autonomy in a consistent manner, SAE International (the Society of Automative Engineers) proposed 6 levels of autonomy in the standard J3016~\cite{av_sae_classification}. There, Level $0$ means no automation and Level $5$ is full automation. From Levels $0-2$, a human driver monitors the driving environment, whereas from Levels $3-5$, the driving system monitors the driving environment. A tabular description of the capabilities associated with different levels of autonomy is provided in Table~\ref{tab:levels}.

\begin{scriptsize}
\begin{table}[hbt]
\centering
  \begin{tabular}{c|c|c|c|c|c}

   \multirow{2}{*}{Level} & \multirow{2}{*}{Automation} & Steering  &  Environment & Fallback & Driving \\
         &             & Cruise  & Monitoring   & Control  & Modes  \\ \hline \hline
   0     & None        & \cellcolor{blue!25}H		& \cellcolor{blue!25}H		& \cellcolor{blue!25}H    	& N/A   \\
   1     & Supportive  & \cellcolor{blue!25}H,S 	& \cellcolor{blue!25}H		& \cellcolor{blue!25}H    	& Some  \\
   2     & Partial     & S		& \cellcolor{blue!25}H		& \cellcolor{blue!25}H    	& Some  \\ \hline
   3     & Conditional & S    	& S		& \cellcolor{blue!25}H    	& Some  \\
   4     & High        & S		& S		& \cellcolor{blue!25}H   	& Some  \\
   5     & Full        & S		& S		& S   	& All   \\ \\

  \end{tabular}
  \caption{Autonomous Vehicles: Levels of Autonomy ($S$ - System, $H$ - Human)}
  \label{tab:levels}
\end{table}
\end{scriptsize}

Understandably, with higher degree of autonomy, the security risks are also escalated. From Level $3$ onwards, it becomes necessary that the car is fitted with an increased number of sensing and communicating devices in order to be "self-aware". In the following discussions, unless specifically mentioned, we define Autonomous Vehicles (AV) as modern vehicles with autonomy features at Level $3$ and above as specified in Table~\ref{tab:levels}.

\subsection*{Related Literature}
\begin{figure}[hbt]
    \centering
    \includegraphics[width=70mm]{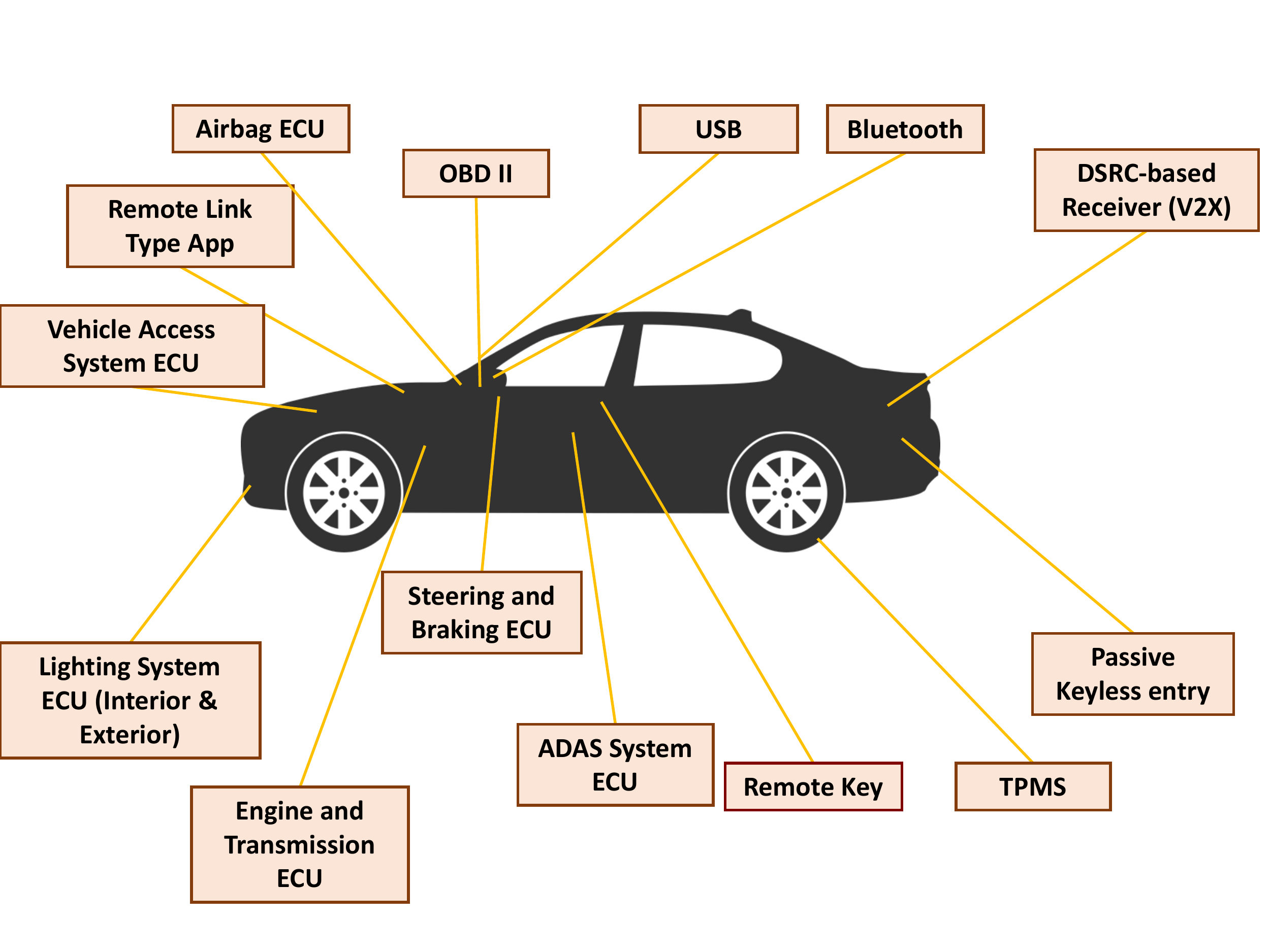}
    \caption{Potential Attack Sources and Surfaces in Cars~\cite{intel_asrw_2016}}
    \label{fig:attack-source-surfaces}
\end{figure}

A connected AV is subjected to cyber attacks through its various network interfaces to the public network infrastructure as well as its direct exposure to the open physical environment. An attack surface of a system is the sum of the different attack vectors, that is the different points where attackers can make attempts to inject data to or extract data from the system in order to compromise the security control of the AV. Figure~\ref{fig:attack-source-surfaces} depicts the typical attack surfaces of an AV, such as the ones identified by the authors in~\cite{intel_asrw_2016}, and potential attack sources through which AV security can be threatened. It can be observed that the attack sources are typically external agent/event or even an internal component with malicious intent that attempts to compromise the expected autonomy functionality of the AV. For example, the Bluetooth interface of the AV shown in Figure~\ref{fig:attack-source-surfaces} can be considered a potential attack surface that can be compromised through plugging malicious devices (attack source) into this communication channel.

There has been a significant body of work on the perceived threats, and the related countermeasures, of AV security~\cite{Lemke_book_2010}\cite{Wolf04securityin}\cite{Checkoway_Automotive_Attack_2011}. In this article, we argue that these threats can be categorized according to the generic attack models of Cyber-Physical Systems (CPS), i.e., originating from the attack surface of an exposed CPS component. In the next section, we briefly review the security issues of CPS as well as the security of AV as a specific kind of CPS; and particularly emphasise on attacks that have been demonstrated in practical setting.

\section{Security Issues of Autonomous Vehicle}
In this section, we first review the security issues of CPS before pinpointing to the key considerations and new challenges that are specific to AV security.

\subsection{CPS Security}
\begin{figure}[h]
    \centering
    \includegraphics[width=110mm]{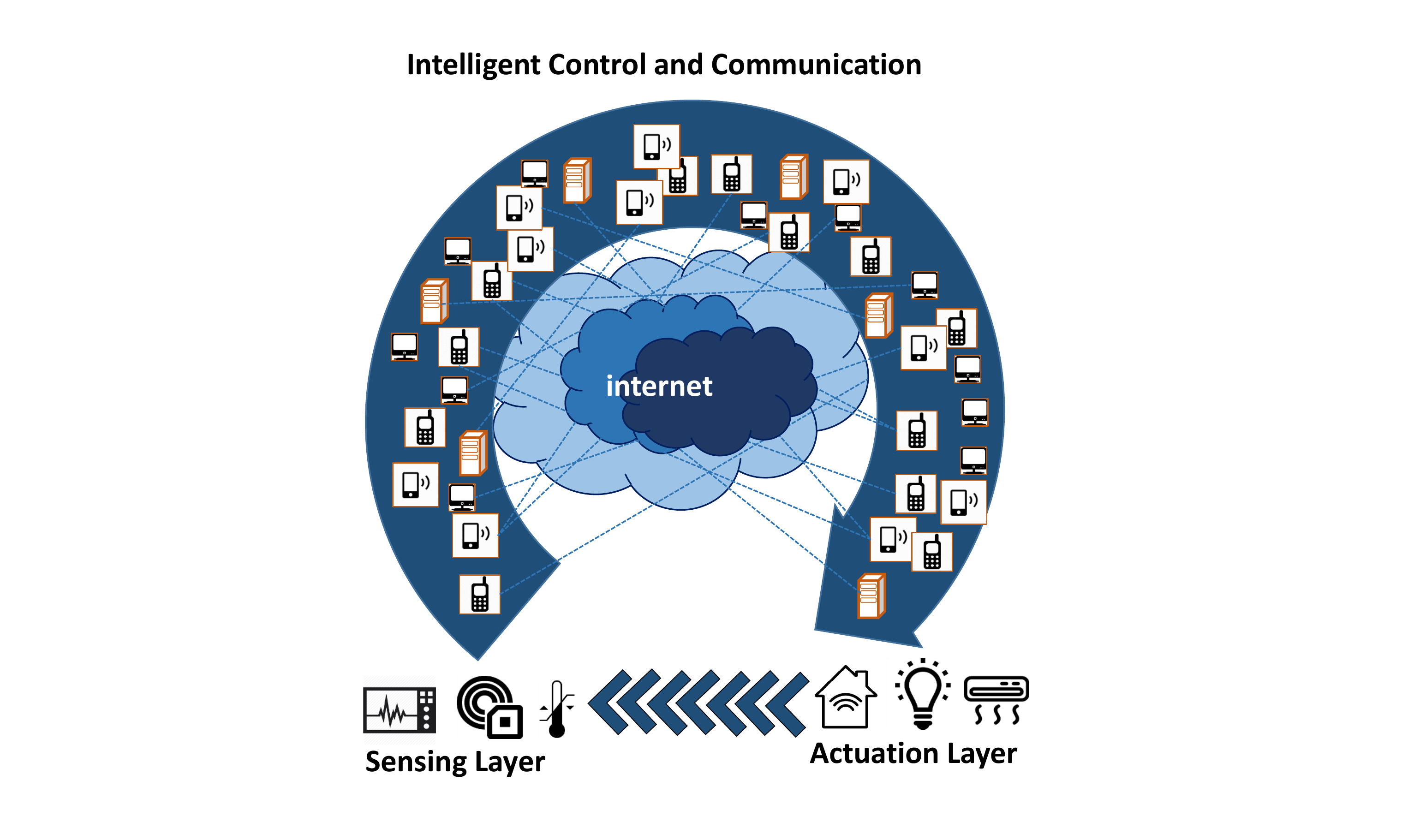}
    \caption{Interactions between sensor layer and actuator layer}
    \label{fig:intelligentC2}
\end{figure}

An autonomous vehicle can be considered a specific type of Cyber-Physical System (CPS) and also a kind of Internet-of-Things (IoT) system. Cyber-Physical-Systems are complex, heterogeneous distributed systems, typically consisting of a large number of sensors and actuators connected to a pool of computing nodes. With the fusion of sensors, computing nodes, and actuators, which are connected through various means of communications, CPS aim to perceive and understand changes in the physical environment, analyze the impacts of such changes to the operation of the CPS, and make intelligent decisions to respond to the changes by issuing commands to control physical objects in the system; thereby influencing the physical environment in an autonomous way~\cite{6096958_Survey}. As illustrated in Fig.~\ref{fig:intelligentC2}, the connections between actuation and sensing through the physical environment, and between sensors and actuators through one or multiple (distributed) computing or intelligent control node(s), form a feedback loop which aims at achieving a desired objective or steady state. As such, a CPS acts either with full autonomy or at least provides support for a human-in-the-loop mechanism as part of some semi-autonomous control functions. This distributed closed-loop process allows CPS to remotely influence, manage, automate and control many industrial operations.

Due to the operational nature of CPS in most industrial control processes, CPS are also known as Operational Technology Systems (OT Systems) \cite{Burg_Anupam_Lam}\cite{kwokyan_security_paloalto}.

The massive adoption of Internet-enabled devices (i.e., IP-enabled sensors and actuators) in CPS systems has thereby blurred the boundary between CPS and Internet-of-Things (IoT). The concept of IoT stems from connected smart devices~\cite{weiser_iot_99}, which may or may not be interacting with physical objects. Hence, there are application scenarios in the classical OT domain that can be conveniently classified both as an IoT system and a CPS system, e.g. distributed set of sensor nodes to monitor and control the energy usage of a manufacturing plant. Prominent examples of CPS and IoT systems, and their corresponding applications~\cite{6853346_Survey} include, among others, autonomous vehicles, which is the focus of this work.

\begin{figure}[hbt]
    \centering
    \includegraphics[width=60mm]{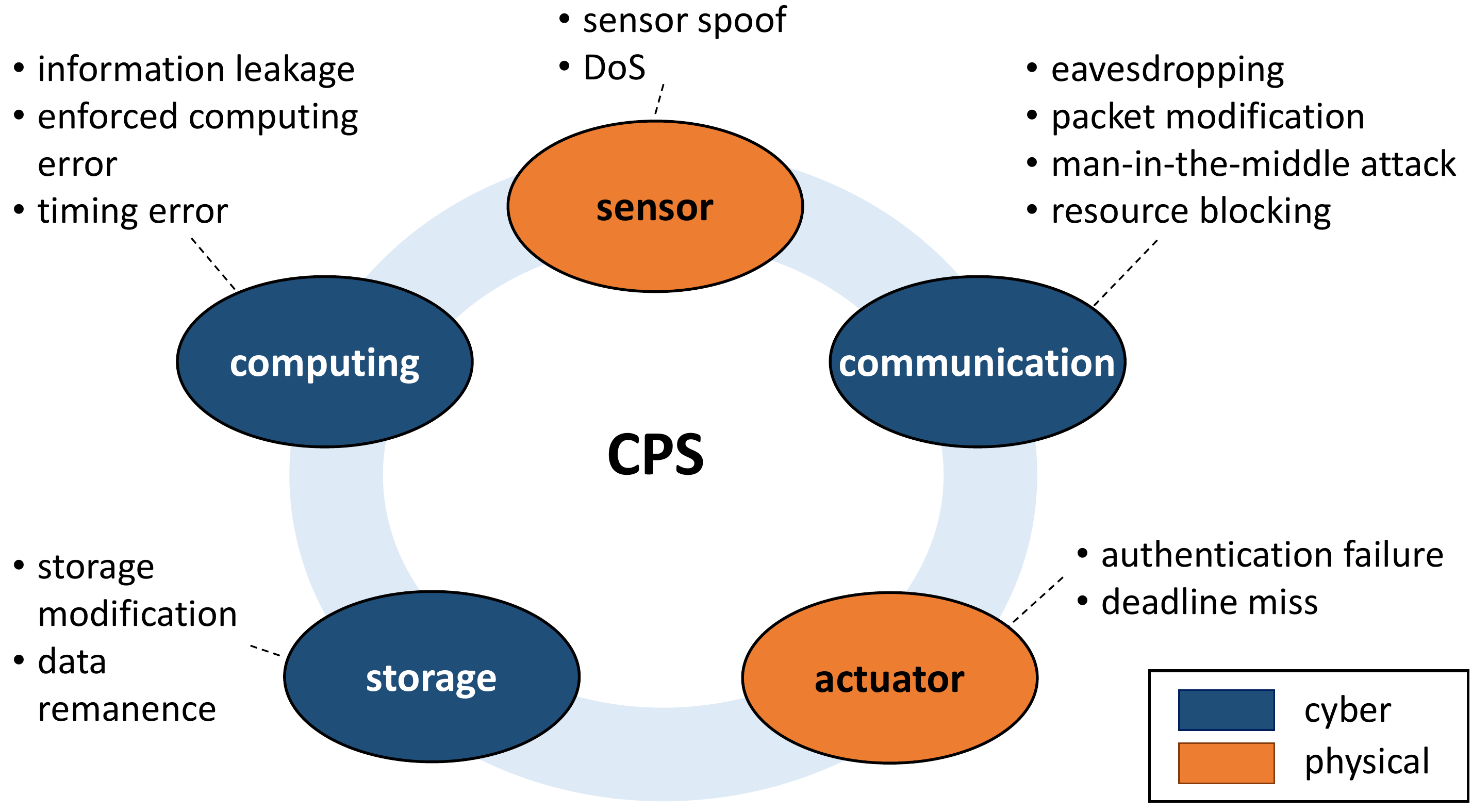}
    \caption{CPS Attacks: Generic Model}
    \label{fig:cps-attacks}
\end{figure}

As such, attack techniques against a diversity of OT systems have similarities and can be classified as attacks on different CPS components, e.g., communication, storage, actuator, sensor and computing nodes. A few such attacks are shown in the Fig.~\ref{fig:cps-attacks}.

However, this na\"ive and isolated analysis of attacks, in the context of a specific CPS, and the adoption of the corresponding countermeasures, are grossly inadequate and misleading for several reasons.

\begin{itemize}
\item These generic attack studies tend to ignore the security objectives of the CPS, which aim to strike a balance between risks, cost and convenience through the adoption of a hybrid of security control measures. Thus, a seemingly insecure mechanism may be operationally acceptable due to the fact that it is operating within a controlled environment created by other security mechanisms of the system.

\item Depending on the prevailing OT security practices, as well as the assumed adversarial model, it might be unnecessary to account for certain vulnerabilities.

\item The generalization of attacks across all CPS typically ignores the roles of Roots of Trust (RoT) and security perimeter modeling, which are the basis of many security-by-design approaches.
\end{itemize}

In principle, security-by-design of a CPS is a holistic process which is viewed as a systems engineering discipline~\cite{nist_sp800_160}. Addressing specific attacks in an isolated and ad-hoc manner cannot help much in security design practices. For these reasons, this paper refers to the above classification and enumeration of attacks, as done predominantly in the current literature, as \textit{generic attack} studies in that they tend to study localized attacks in a generic setting of CPS. This trend of generic attack studies is exacerbated by the fact that there is no well-defined security standard that aligns with the road-safety standards for the AV.

In contrast, we emphasise on the security-by-design of AV as a system, indeed a cyber-physical system. In the context of the specific vulnerabilities of AV with different degrees of autonomy, the technological challenges to safeguard AV are derived directly from the underlying safety objectives.

\begin{figure}[h]
    \centering
    \includegraphics[width=90mm]{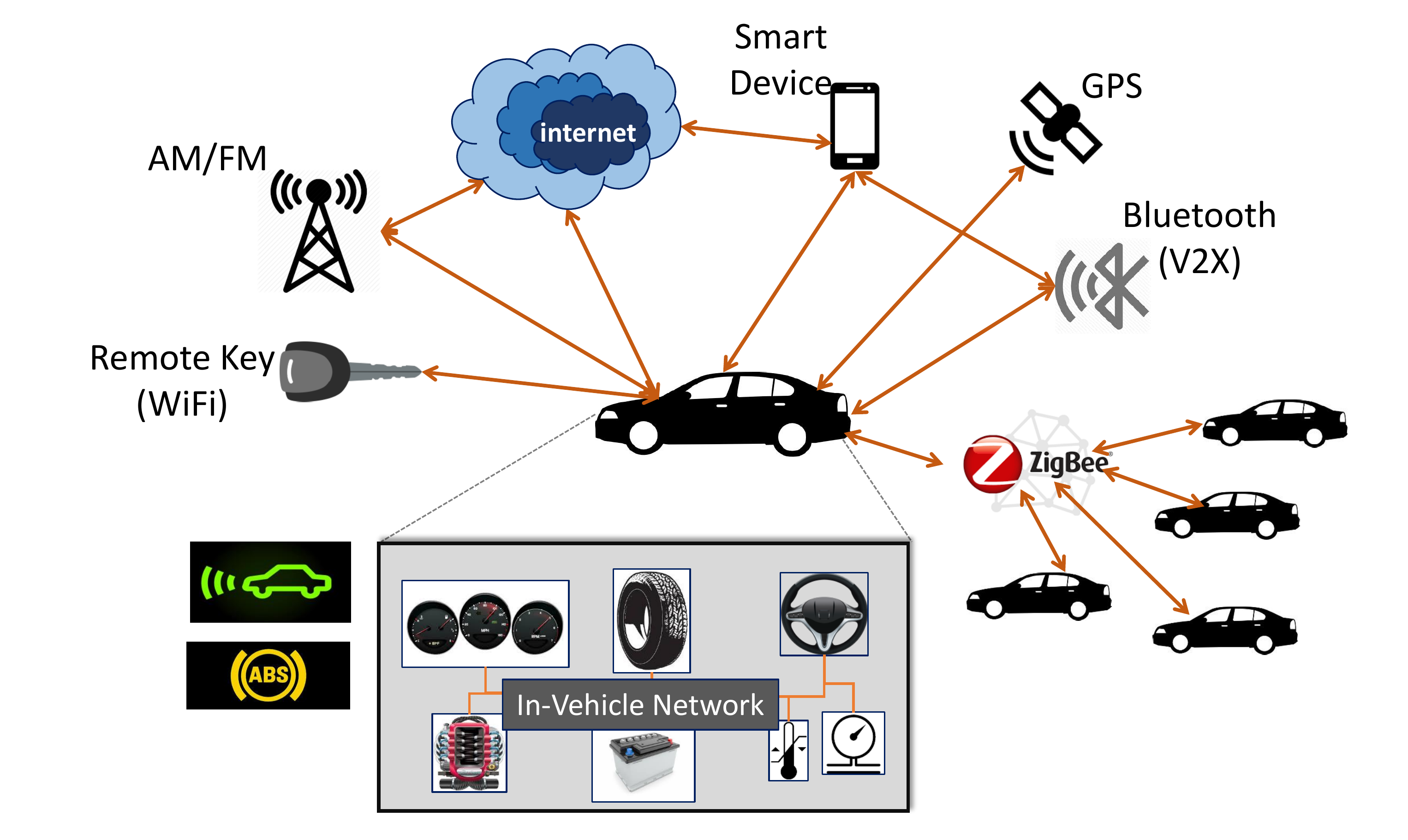}
    \caption{Exemplary Networked Vehicle}
    \label{fig:car-security}
\end{figure}

\subsection{AV Security Issues}
The networking capabilities of a typical AV is depicted in Fig.~\ref{fig:car-security}. The car is connected to the outside world with vehicle-to-infrastructure (V2I) and vehicle-to-vehicle (V2V) communication links. The V2I and V2V links provide the AV with, for example, traffic-status information from traffic management infrastructure or navigation-related information received from other AVs on the road. These connection interfaces represent attack surfaces that the adversaries will aim to exploit in order to obtain unauthorized access to the AV. Thus, it is of paramount importance that these V2I and V2V connections are mutually authenticated and the payload suitably protected from unauthorized disclosure and unauthorized modification.


On the other hand, the internal of an AV is installed with a multitude of independent and overlapping CPS systems, such as adaptive cruise control, anti-lock braking system, assistive parking system. Depending on the level of autonomy, more functionalities are helmed by the internal CPS controllers, with varying degree of human intervention. These CPS systems are supported by actuators and sensors, such as tyre pressure sensor, crankshaft position sensor, light sensor and obstacle sensors. Today, intra-vehicle communications for supporting such CPS controls happen mostly through wired connections known as the Control Area Network (CAN). The intra-vehicle communication protocols typically follow the CAN bus standard or the more recent FlexRay standard.

The security design of AV requires that intra-vehicle networks~\cite{tuohy_intravnetwork_15}\cite{faezipour_intravnetwork_12} are rigourously protected and that access from outside of the vehicle, if applicable (e.g. for maintenance and services of the AV), be strictly controlled. It is evident from previous generic attack studies of AV security that, without proper security designs, even communications through these intra-vehicle wireline standards are potentially susceptible to security breaches~\cite{Checkoway_Automotive_Attack_2011}\cite{flexray_attack}. Specific studies of the intra-vehicle network security has been undertaken in~\cite{Wolf04securityin}, which is, however, decoupled from the holistic view of the security-by-design principles.

In addition, connections to the internet enable the AV to transmit operational data to the car manufacturer. As a critical system from the perspective of autonomy functions, it is crucial that the communication link between the AV and its manufacturer be authenticated and the payload be appropriately protected in accordance with the nature of the operational data being transmitted from the AV to the manufacturer.

\section{Security-by-Design}
Security designs of business or industrial systems always start from the security objectives of the systems which have to take into consideration the management and people aspects of the systems~\cite{nist_sp800_64}. Indeed, in real world situations, security requirements/objectives always start from people e.g. defining business objectives of a system by the business owner, making risk-taking decisions is part and parcel of any business venture. Security is meaningful only when people understands what is at stake. In this connection, the consideration is not only in terms of the cost of security attacks but also the cost of implementing security controls as well as the opportunity cost of limiting business operations due to risk-averse designs. As such, system security is modeled as a socio-technical system in which technical design decisions are heavily influenced by people's factors.

In the context of AV, the prime security objective is inevitably the resilience of safety features of the autonomy functions. As human lives are at stake, it is of utmost importance that safety of AV are not compromised in face of attacks on the automony functions. Ultimately, AV demands the highest level of security measures to guard against attacks of safety-related features. This is especially so for CPS responsible for collision detection, cruise control and anti-lock braking etc., because any compromise of such features is likely to result in injuries and even loss of life.

When studying the cybersecurity of socio-technical systems, it is useful to identify the steps and processes where people are involved in the security consideration of the system life-cycle. The people/social/environment factors are typically some of the major sources of vulnerabilities, which vary from system to system. A more detailed discussion on socio-technical issues related to security is available at~\cite{dagstuhl_2014}\cite{dagstuhl_2016}.

\begin{figure}[hbt]
    \centering
    \includegraphics[width=65mm]{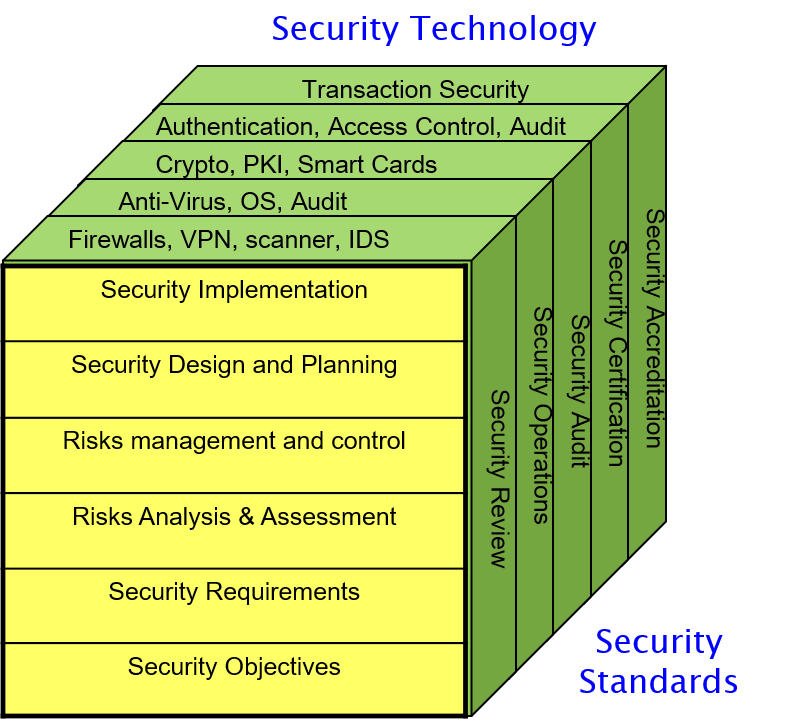}
    \caption{Layers of Security Management}
    \label{fig:cps-security-management}
\end{figure}

A general overview of the security management framework for a socio-technical system is presented in Fig.~\ref{fig:cps-security-management}. In this paper, we discuss the security-by-design of AV as a CPS, and in a holistic manner, by identifying and addressing the security objectives within this socio-technical framework. In the next subsection, important concepts that form the basis of security design of AV systems are reviewed.

\subsection{Tenets of Security}
When designing security for complex systems, it typically involves the following key steps: (1) identify the security objectives and requirements of the system; (2) assess the value and sensitivity of the system to be protected; (3) define security policy in accordance with the security requirements; (4) estimate the capabilities of the adversaries; (5) design control features that commensurate with the sensitivity of the system and the risks it is exposed to.

Traditionally, standard security policies are guided by the generic Confidentiality, Integrity and Availability (also known as the CIA triad) requirements at the message/data level. These policies are reinforced with application-specific security requirements and are eventually implemented with a combination of physical and logical control measures including cryptographic primitives and security protocols.

An alternative approach towards security design is to model the potential threats at a system level. For example, Microsoft proposed a threat classification model based on the following six categories, also termed as the STRIDE threat model~\cite{ms_stride}.
\begin{itemize}
\item \textbf{S}poofing of user identity.
\item \textbf{T}ampering of stored or communicated data.
\item \textbf{R}epudiation, i.e., denying of actions performed where other users cannot prove otherwise.
\item \textbf{I}nformation disclosure or breach of confidentiality.
\item \textbf{D}enial of Service (DoS) which makes a network/server unavailable.
\item \textbf{E}levation of privilege for an user to perform unauthorized actions.
\end{itemize}	

In modern CPS and IoT systems, this threat model is discussed together with the notion of attack surfaces. As explained before, an attack surface provides an entry point for an attacker to gain control of or exfiltrate information from the target system. A detailed study of automotive attack surfaces is presented in~\cite{Checkoway_Automotive_Attack_2011}.

\subsection{Adversarial Model}
Once the attack surfaces of a system are identified, security designers will need to estimate the likelihood that attacks may happen in real and operational situations. This will mainly depend on (1) the capabilities of the adversaries as well as (2) their interests in investing resources to perform the attacks. The former is typically analyzed by means of some adversarial models while the latter depends on the value and sensitivity of the system to both the system owner and the attacker.

Adversarial model is a formal definition of the attackers' capabilities.
Dolev and Yao defined the adversarial model in~\cite{dolev_yao_it_83} for proving the properties in an interacting cryptographic system. The adversary, in this model, is capable of hearing, intercepting, initiating and synthesizing any message. However, such a $strong$ adversarial model is considered unrealistic; instead $practical$ adversarial models, validated by practical experiments, are used often. Adversarial models have also been used later for network security, e.g., for the security analysis of Onion Ring Routing, which permitted traffic-analysis-resistant and anonymous Internet connections~\cite{Syverson_onion_routing_2001}. Adversarial models can also be used for analyzing privacy guarantees, e.g., as has been done in~\cite{heiber_privacy_2005}, which defined an adversary capable of intercepting the messages and identifying the location of the context-aware system.

In the following, we define different parameters and possible adversarial models for analyzing CPS security. Since the adversarial model is closely tied to the application domain, we use AV as a representative scenario here.

First, the \textit{attack objective}. For a given CPS, the attacker attempts to breach one or more of the security objectives. For example, integrity breaches of an AV would mean that the attacker is potentially able to take over control of the vehicle. As another example, by being able to control communicatons between the AV and any third-party system, the attacker is capable of violating the  confidentiality and integrity of the communication channel, hence can potentially manipulate the safety related control functions of the AV.

Second, the \textit{communication capability}. Adversarial model is mainly defined by the overall communication capability of the attackers, e.g., to intercept messages transmitted over different communication channels of a CPS, which determines what kind of messages are available to them for analysis and their ability to inject maliciously fabricated messages to the system. In the context of an AV, this could be commands sent through the internal CAN bus of the vehicle, or could be messages transmitted as part of some V2X communication.

Third, the \textit{computing capability} of the adversaries. Even after gaining access to the CPS, through breaching the trusted network, the adversary needs to execute tasks to gain control or damage the CPS. These typically require cryptanalysis of cryptograms, decoding of application protocol messages and reverse-engineering security-critical parameters. The attackers will need to carry out these tasks online or offline in an efficient manner, often requiring considerable computing power.

In the context of AV security, there is no well-defined adversarial model, hence making the work of a security designer harder. Some recent works in the open literature defined adversarial deep learning~\cite{berkeley_deep_drive} that can undermine the autonomous driving controls. However, this is not the kind of adversarial model for analyzing security of the overall AV operations, and is outside the scope of this discussion.

\subsection{Trust Model and Security}
As mentioned, security objectives are typically fulfilled by cryptography-enabled control mechanisms which aim to achieve confidentiality, integrity and authenticity, at least, at the data and message level. This leads to the requirement for key management, which in turn are based on some Trust Model. For example, a Trusted Third Party (TPP) played by an Authentication Server for symmetric key systems or a Certification Authority of public key systems. Similarly, the use of Web of Trust as in the case of PGP, or Distributed Trust adopted by Blockchain, are examples of Trust Models. In the latter, due to the absence of a TPP, one may adhere to the use of consensus protocols and decision mechanisms like the de-centralized trust model of Blockchain.

Trust Model is fundamental to all practical security designs because it lays the security foundation of who one can trust, which in turn allows users to determine what can be trusted. A good security design always starts from a realistic and practical trust model, with other cryptographic mechanisms and security protocols being established and proven on the basis of the trust model.

To illustrate, if the security of a CPS is dependent on information/instruction received from another party, whose trustworthiness is dubious, then the security of the CPS is ill-based. Unfortunately, from experiences, this kind of situation happened very often in practice, most likely, due to the retro-fitting approach to security of a lot of existing CPS. For example, a lot of sensitive online transactions have been hacked because the underlying key management mechanisms, which are assumed to be trusted, could be based on unverified open source codes wherein security loopholes are commonplaces. \textit{If the approach of security-by-design is adopted, the underlying trust model will need to be explicitly defined and the basis of such trust will have to be verified and assured.}

\subsection{Trust Infrastructure and Security Mechanisms}
In the security design of AV, the communication connectivity infrastructure is of prime importance; besides, it is imperative to support security administration of trustworthy IoT devices such as key, identity and privilege managements of these devices. In a startling difference from wired networks, for wireless and mobile ad-hoc networks, it often admits new members, and therefore needs to frequently establish secure communication channels. In the following, the distinguishing features of a secure and trustworthy network management infrastructure are presented:

\begin{itemize}
\item \textit{Infrastructure for trust management}: A prime use case scenario for IoT devices are ad-hoc sensor networks, which find applications in V2V and V2I communications, for example. In such networks, control mechanisms for admitting new nodes and detecting malicious nodes~\cite{butun_intrusion_detection_14} are important prerequisites for maintaining security policies intact. There have been ample studies on the key management protocols for wireless sensor networks, e.g., via key pre-distribution~\cite{bechkit_key_predistribution}, identity-based encryption~\cite{chu_ibe_wsn}\cite{zhang_ibe_vehicular}, certification authorities and key exchange protocols. In general, these studies fall under the general theme of trust management~\cite{iot_asokan_16}, which is particularly challenging for low-end devices due to the performance overhead that a secure key storage or dynamic code attestation incurs.

\item \textit{Secure routing protocol}: IoT systems rely critically on static/dynamic routing protocols, which may be subjected to diverse forms of attacks~\cite{karlof_secure_routing}. Typical countermeasures for routing protocol attacks depend on, firstly, a trusted base station that enables authentication and encryption; secondly, multipath routing and, thirdly, secure geographic routing protocols~\cite{garcia_otero_2010}.

\item \textit{Heterogeneous network integration}: IoT networks are usually associated with a heterogeneous mix of wireless communication systems, each of which comes with its own security protocols \cite{Burg_Anupam_Lam}. Their interoperability may require conversion of data formats, which is difficult to undertake without partial knowledge of the message payload. Furthermore, the possible presence of, and often undetected, diverse information channels remains a constant threat~\cite{airhopper} that needs to be addressed by secure and trustworthy network management means.

\item \textit{Secured Resources}: The computing and storage blocks of a CPS can be considered secured, e.g., as in a Trusted Execution Environment (TEE). The resources under this category need to be assured by only accepting authenticated and signed boot process. Secure storages can be designed by preventing known attacks like, data remanence; and further opting for encrypted and authenticated data storage.

\item \textit{Trusted Identification}: For a CPS resource to be included in the trusted network, it must participate in a robust identification protocol, or be certified by a centralized certification authority. This works in similar principles as in wireless sensor networks, where trust management~\cite{iot_asokan_16} becomes challenging for low-end devices. For such a resource, trust anchoring can be done using Physically Unclonable Functions (PUFs). Typical adversarial models against these are hardware Trojan elements, introduced in the manufacturing/repairing process.
\end{itemize}

\section{Security-by-Design for Autonomous Vehicles}
In this section, we describe the security assumptions, requirements, threats and control measures of AV by adopting the security-by-design approach. Specifically, we describe the key steps of security-by-design in the context of AV by viewing it as a socio-technical system. As explained, when designing security for complex systems such as the AV, it typically involves the following key steps: (1) identify the security objectives and requirements of the system; (2) assess the value and sensitivity of the system to be protected; (3) define security policy in accordance with the security requirements; (4) estimate the capabilities of the adversaries; (5) design control features that commensurate with the sensitivity of the system and the risks it is exposed to.

In order to commence the security-by-design process within a sound framework, it is important to establish the operation model of the AV so that security objectives and requirements may be analyzed in a holistic and systematic manner.

\subsection{AV Operation Model}\label{ssec:av_model}
For simplicity, the following \textit{basic operation model} of an AV is assumed.
\begin{itemize}
\item \emph{Communication:} AV periodically sends operation logs to the manufacturer to allow life cycle management and maintenance.
\item \emph{Communication:} AV has wireless or wired interfaces to support firmware/software update/upgrade at maintenance and service workshop.
\item \emph{Communication:} AV supports communication with some traffic management system infrastructure for traffic flow control as well as remote control in case of emergency.

\item \emph{Sensing:} AV is equipped with a variety of sensors to sense the physical environment and detect collisions.
\item \emph{Decision:} AV has a number of navigation-related control functions that allow Level-5 autonomy, i.e., it requires real-time updating of travel routes, enables intelligent route planning and automatic steering in accordance with road conditions.
\item \emph{Decision:} AV has a number of safety-related control functions that allow Level-5 autonomy, i.e., it enables automatic steering, speed regulation and braking in accordance with road conditions.
\end{itemize}

\subsection{Security Objectives and Requirements of AV}\label{ssec:sec_obj}
Based on this operation model, we identify the basic security objectives of an AV as follows:
\begin{itemize}
\item \emph{Integrity} of remote control functions of the AV (possibly as an emergency operation from the traffic management system) so that no attacker is able to take over control of the AV by tampering the remote control system.
\item \emph{Integrity} of the sensor systems so that navigation and safety related control features will not be interfered by attackers through tampering the sensor data.
\item \emph{Integrity} of the safety-related control operations such as braking and speed control are performed in accordance with the sensed road conditions or from remote control instructions.
\item \emph{Integrity} of the navigation-related control operations such as steering, braking and speed control are performed in accordance with the sensed road conditions or from pre-programming route path.
\item \emph{Confidentiality} of communications between AV and traffic management system so that safety-related control parameters will not be disclosed to unauthorized parties who may potential develop further exploits to the AV.
\item \emph{Confidentiality} and \emph{integrity} of communications between AV and its manufacturer so that robustness of the life cycle management and maintenance of the AV can be assured.
\item \emph{Integrity} and \emph{authenticity} of communications between AV and maintenance workshop so that software patches and updates to the AV can be performed with high assurance.
\item \emph{Confidentiality} of cryptographic keying materials stored inside the AV are ensured so that attackers cannot bypass higher level security mechanisms by siphoning the cryptographic keys.
\end{itemize}

\subsection{Safety Standards for AV}
AV should be modeled as a socio-technical system where safety is of utmost importance because human lives are at stake. For AV, safety-criticality directly leads to the criticality of cybersecurity, and not the other way around. Hence, the AV security objectives are to be based on, and derived from, the relevant AV safety standards. The current safety standards for AV are presented in the following.

\begin{itemize}
\item ISO 26262~\cite{iso_26262}: This standard is derived from IEC 61508~\cite{iec_61508}, which was developed for all electrical/electronic safety-related systems. ISO 26262 is specifically targeted for automotive safety. This standard provides a safety lifecycle throughout the phases like management, development, production, operation, service, and decommissioning. ISO 26262 also defines the Automotive Safety Integrity Level (ASIL). ASIL includes Severity classification (S0 -- S3), Exposure classification (E0 -- E4) and Controllability classification (C0 -- C3) to quantify the severity of an injury, probability of occurrence and controllability of the situation, respectively. ASIL is expressed as follows.

\begin{equation*}
ASIL = Severity \times Exposure \times Controllability
\end{equation*}

where the higher level of ASIL indicates a more grievous situation. In the context of AV, it can be noted that the controllability level is extremely high for level 3 upwards. To assess the ASIL, one can adopt techniques such as, Hazard Analysis and Risk Assessment (HARA), Fault Tree Analysis (FTA), and Failure Mode and Effects Analysis (FMEA).

\item SAE J3061\cite{av_sae_classification}: Recognising the specific need for a standard in the wake of cyber security incidents for automotive, the Society of Automotive Engineers (SAE) decided to move together with ISO to define the standard J3061. While it is based on the ISO 26262, it identifies the growing threat landscape and tries to establish the awareness and a common terminology across the AV supply chain. It establishes the terminology of threat (malicious attacker), vulnerability (unguarded gateway) and risk (likelihood of attack). Most importantly, J3061 delineates the scope of cybersecurity by stating that -- cybersecurity-critical system are not necessarily safety-critical, however, the reverse is true. It also emphasizes the distinction between system safety (fault/accident) and system cybersecurity (purposeful attack).
\end{itemize}

\begin{figure}[hbt]
    \centering
    \includegraphics[width=90mm]{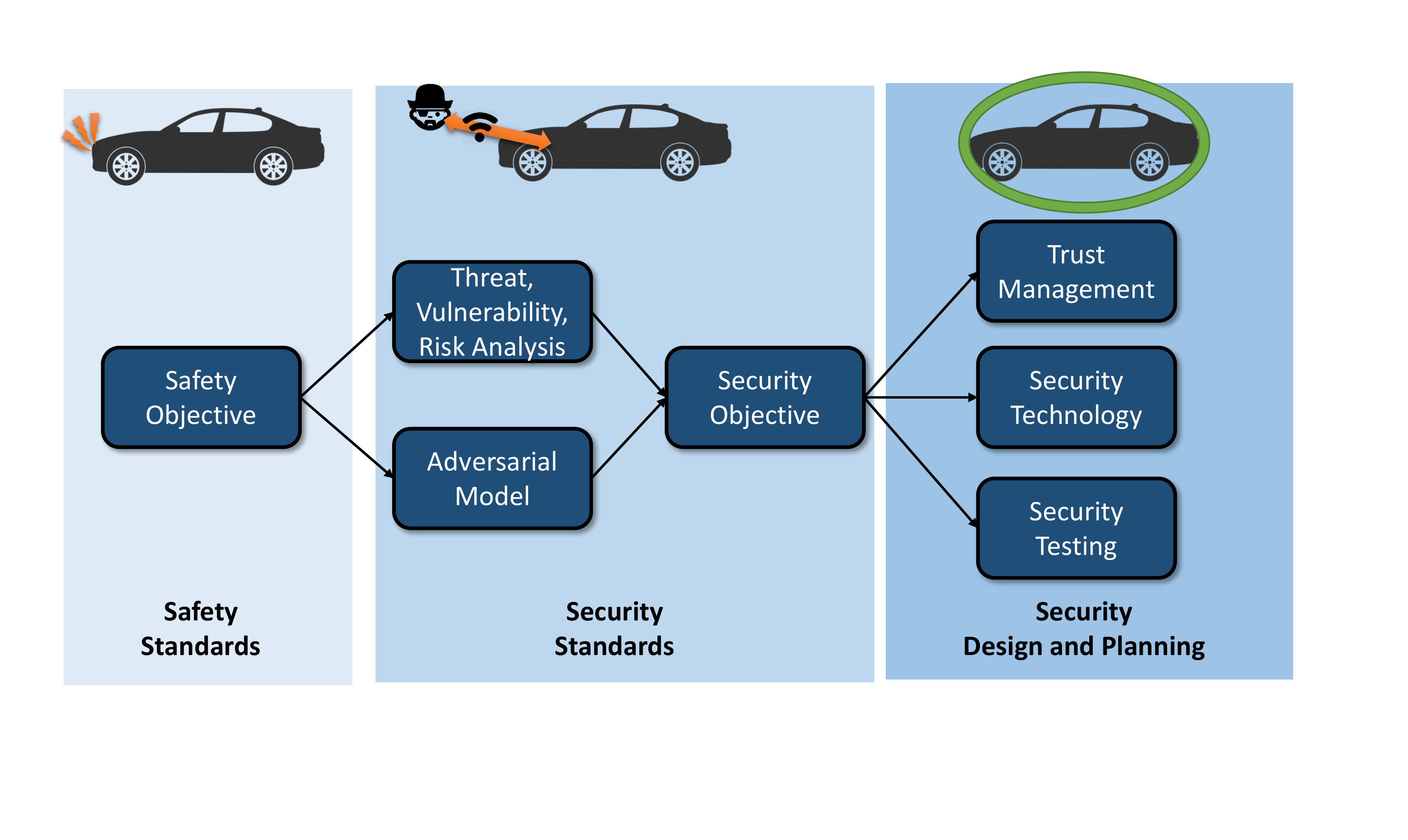}
    \caption{AV Security Design Flow}
    \label{fig:av-security-safety}
\end{figure}

Figure~\ref{fig:av-security-safety} illustrates our approach for analyzing and identifying the AV security objectives and requirements on the basis of the relevant AV safety and security policies and standards.

\subsection{Adversarial Models for AV Security}\label{ssec:adversary}
Given the security objectives and requirements as well as the system security model for AV, we now consider the assumed capabilities of the attacks; that is the adversarial models for AV security. As discussed, the purpose of this effort is to estimate the capabilities of the adversaries so that control features can be designed to commensurate with the sensitivity of the system and the risks it is exposed to. The following adversarial capabilities are defined for AV, which are applied for the security analysis for this manuscript.

\textit{
\begin{Properties}
  \item An adversary is capable of intercepting and tampering all inter-vehicle and intra-vehicle communication.
  \item An adversary is capable of introducing malicious nodes in the inter-vehicle and intra-vehicle communication network.
\end{Properties}
}

These adversarial models are practical and have been demonstrated in the context of several attacks.

\subsection{System Security Model for AV}\label{ssec:av_sys_sec}
Though the aforementioned standards can act as the guidelines to begin with, yet there is a lack of consistency in the approaches for identifying attack surfaces, threat identification and risk assessment. Consequently, there is no single standard applicable to the AV security as a CPS given its complex overlap across multiple technology domains like wireless communication, electronics~\cite{iso_26262}, mechanical systems and software development practices~\cite{iso_27034}.

\begin{figure}[hbt]
    \centering
    \includegraphics[width=90mm]{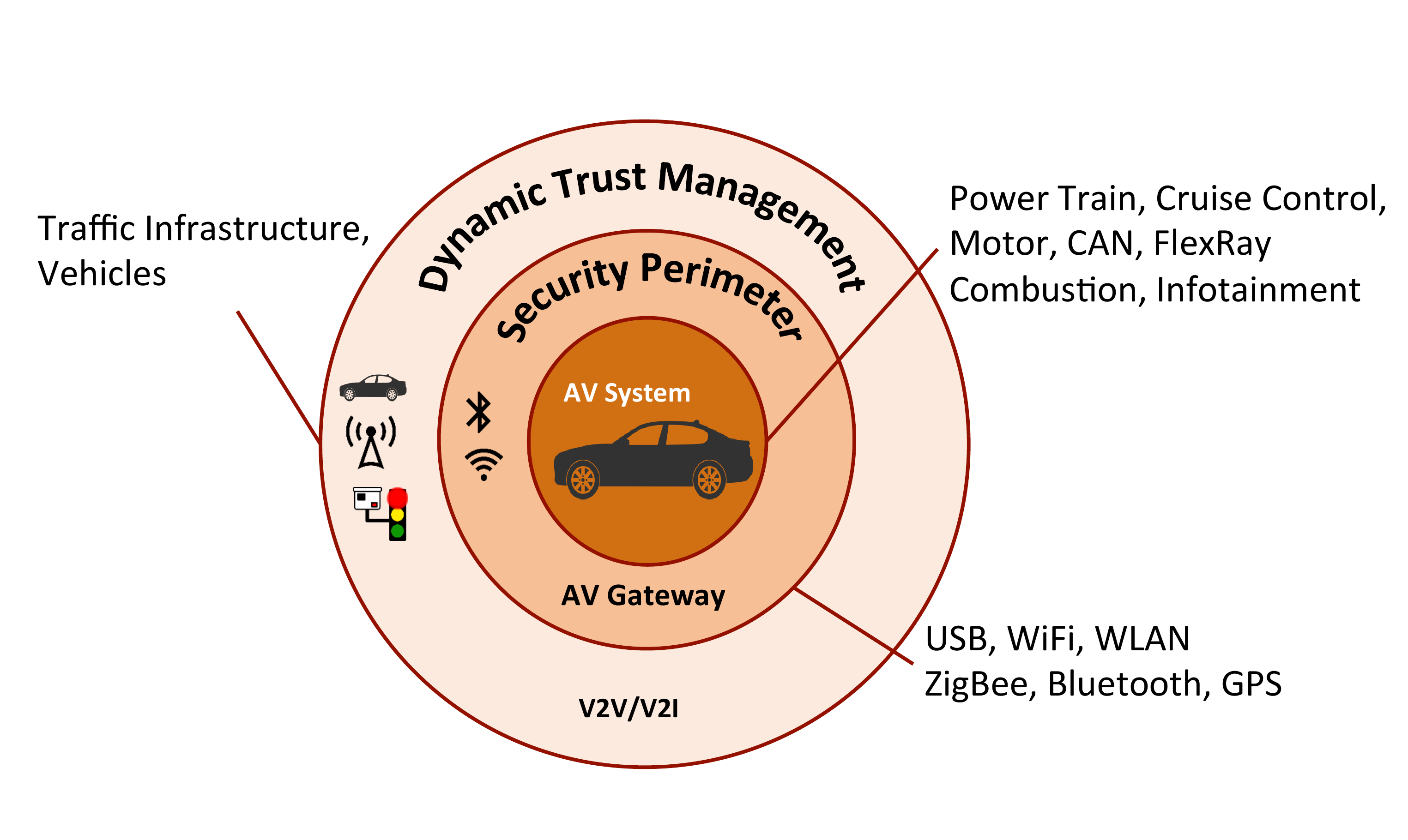}
    \caption{AV Security by Design}
    \label{fig:av-security-by-design}
\end{figure}

Figure~\ref{fig:av-security-by-design} presents a high-level model for security-by-design of AV. In this model, we define the attack surfaces of the AV into three layers: (1) the core layer which is defined by the physical enclosure of the AV; (2) the interface layer, or AV gateway layer, which is characterised by the collection of connectivity interfaces between the AV and the external world; (3) the infrastructure layer which is composed of all the infrastructure and backend modules which are trusted by and connected to the AV. The security issues are clearly distributed over these three layers, first, the AV system; second, the AV gateway and third, the V2V/V2I communications.

The core layer, being defined by the physical enclosure of the AV, enforces its perimeter through physical security. The AV gateway layer consists of diverse forms of communication links, including Bluetooth, WiFi, ZigBee that helps establish and maintain communication with external vehicles and infrastructures. This layer defines a network security perimeter, which is enforced by network access control mechanisms to guard against unauthorized access to the internal functionalities of the AV. This layer relies on the assumed physical security of the core layer so that any damage done to this vehicle within this layer through physical means are not  considered an AV security hazard due to cyber attacks.

The outermost layer includes all external systems that are interacting with the AV, which includes vehicles, traffic infrastructures and cloud-based navigation service provider, for example. Note that, not all of these external systems could be upfront categorized as a trusted party. Therefore, a dynamic trust management approach has to be undertaken at this layer. A threat arises when a trusted component communicates malicious packets (e.g., malware during software update) or when an untrusted component is able to bypass the secure gateway (e.g., by compromising the vehicle sim card). Note that, the AV system is not necessarily falling prey to an attacker if the security perimeter is breached. For example, Audi A8 maintains a network layout~\cite{audi_security}, where the wireless infrastructure is kept away from the internal AV network via a secure gateway. Thus, a cybersecurity incident may occur but, safety will not be compromised provided the safety related features are not accessible from the wireless gateway.

\section{AV Security: Implementation, Operation and Management}
Typically, the design of security controls starts with the establishment of a security perimeter, followed by a root-of-trust (ROT) definition, and the design of trust infrastructure providing the basis for trust management discussed in the following subsections~\ref{ssec:perim} and \ref{ssec:rot}, respectively. This works towards achieving the security objective based on the AV system security model (section~\ref{ssec:av_sys_sec}). Note that, the security objectives are derived from the safety objective, via TVR analysis and the study of adversarial models.

On the basis of the threat environments due to these perimeter and ROT definitions, we can identify potential security vulnerabilities of AV and, through the identified attack surfaces, can design the corresponding countermeasures. These countermeasures will be included in the secure AV design framework. We also suggest how these threats/countermeasures could be used as the parameters for AV security testing/auditing in connection to the cybersecurity analysis techniques.

\subsection{Security Perimeter}\label{ssec:perim}
The purpose of security perimeter is essentially to divide the AV into segregated security domains with different threat environments. This segregation will have direct impact on the validity of the trust model, hence the design of the underlying trust infrastructure as well as the security mechanisms. By establishing and accepting a security perimeter, AV design can simplify the distinction between cybersecurity incidents and physical tampering.

More importantly, the use of security perimeter allows a holistic and systematic approach to security analysis and design for AV. The banking industry, for example, has accumulated vast experiences in the adoption of security perimeter and defence-in-depth in managing security risks in an organized and controlled manner, when their banking systems interface with the open and hostile Cyberspace, in order to provide Internet banking services.

The three layers defined in \ref{ssec:av_sys_sec} serves as a sound basis for defining the security perimeters for AV. A well-designed security architecture for AV should implement sufficient control mechanisms to reinforce the boundary of each layers so that realistic security assumptions can be made in order to allow systematic security analysis and design for the layers above it.

To illustrate the importance of the notion of security perimeter in AV, we particularly note that a few related literature emphasizes the sensor spoofing attack as a potential threat to AV. In our model defined in  \ref{ssec:av_sys_sec}, this will not be considered a relevant or applicable threat, because the sensors are embedded within security perimeter of the core layer. Without open network distribution of the sensors, any attempt to spoof the sensors will require physical intrusion of the AV, e.g., an attacker may simply damage the rearview camera, thus not considered a cyber attack. On the other hand, a threat is recognized cyber attacks only if it is due to compromises of the network access control mechanisms at the AV gateway layer or tampered messages from a trusted, but compromised, component.

\subsection{Establishing Roots of Trust}\label{ssec:rot}
In order to maintain trusted communications with other CPS/IoT devices internal and external to the AV, the ROT must be established. The ROT will serve as the basis for secure key exchange and for key management. Traditionally, a trusted third party (TTP) may be established to enable key exchange and entity authentication. The TTP will plays the role of assuring the association of a cryptographic key with the truthful identity of the key owner. Besides, the AV is also assumed to have a secure environment for storing the secret cryptographic key.

In general, though some tamper-resistant hardware may be deployed for storing cryptographic keys in a physically protected environment, the association between the key and the device identity remains a security challenge. More recently, due to the cost and operational overheads of deploying and administering remote devices such as IoT sensors and actuators, the use of Physically Unclonable Function (PUF) has attracted the attention of IoT security researchers. In essence, PUF allows the device identity to be established directly from the silicon characteristics of the device itself, hence alleviating the device identity adminustration problem which is fundamental to trust management. Hence, ROT can be achieved by either a PUF-based digital authentication and signature generation; or a public-key protocol to share the key; or having pre-distributed set of keys.

\subsubsection*{Trusted Infrastructure}
A trusted infrastructure can be agreed upon during the AV manufacturing, and during the AV life cycle. The AV design house needs to provide regular support to the end users, for secure software patch, for secure boot and prevention of theft. This can be achieved by marking certain parties as trusted. The smart phone industry is a prominent example where a number of comprehensive PKI-based trusted infrastructure have been established for enabling device lifecycle management, system software upgrade, application software update as well as for prevention of device theft.

Furthermore, in this context, AV design can be further secured by ensuring a secure supply-chain management, which is also a common and recurring theme in the supply chain of Integrated Circuits (IC). Specifically, the planting of Trojan hardware in the AV is a major threat to take into account.

\subsubsection*{Trust Management}
For V2V/V2I interaction in a given AV environment, the notion of trust has to be rigourously maintained and the level of trust be dynamically established. For example, V2V communication should be based on a suitable level of trust among the communicating vehicles. Without this, the V2V communication should not be used for supporting critical operations such as safety-related functions in that any malicious (or compromised) participants in the V2V communication will immediately expose a cluster of AVs to a chain-reaction of attacks. Accordingly, the trust in the AV network has to be managed dynamically. On the other hand, for in-vehicle software update, the patch must be obtained from a trusted source. Though such trusted communication may be established on a need-to basis, the notion of trust must be built on some established ROT.

\subsection{Security of Safety Operations}

We now discuss, through exemplary safety scenarios (specifically collision avoidance and navigation), how some of the key security objectives defined in \ref{ssec:sec_obj} can be achieved. For all the example cases we assume the following RoT and security perimeter.

\begin{itemize}
\item Root-of-Trust: RoT in this scenario includes navigation service provider, which updates and assists on the static/dynamic obstacles in the route of the AV. Furthermore, the integrity of the safety/navigation control components need to be ensured.

\item Security Perimeter: We assume it to be AV gateway and the trusted traffic infrastructure for providing navigational services. This security perimeter can be achieved by establishing a secure gateway for the navigation service provider as well as verifying integrity of the commands.
\end{itemize}

For the security of safety operations, we exemplify the principles through few following practical scenarios.

\subsubsection{Collision Avoidance}
The attacker is able to force the AV into collision with another dynamic object (AV, pedestrian) or a static object (road sign).

\textit{Security-by-Design:} While collision avoidance for AV is a recent concern that requires delving into legal complications apart from the technical challenges, this is a well explored topic in the context of vessel~\cite{szlapczynski_2005} and aircraft navigation~\cite{uav_collision_10}. Similar techniques with precise location update and identification of surrounding objects can ensure that the AV does not collide due to its own failure. However, from the safety angle, the AV should have a high assurance mechanism for independently detecting collision, which will have a over-writing privilege to stop the vehicle gracefully in the event that imminent collision is detected. To achieve this, efficient, real-time embedded obstacle sensors will be desirable, since the assurance of such kill switch mechanism is difficult to attain if it is implemented in software. Worse yet, human lives will be at stake if the collision detection mechanism is based on some AI or analytic techniques whose results tend to be probabilistic.

Here, one can ensure collision avoidance by designing an efficient, real-time, \textit{embedded obstacle sensor-actuator system}. The objective of this system will be to override the car safety/navigation switches and to brake the car and stop the engine, immediately upon the detection of an obstacle that might result in a collision.

To aid this system, the security objective is to ensure - integrity of safety and navigation related control operations. The specific security measures required are as following.

\begin{itemize}
\item \emph{Integrity} of the sensor systems so that navigation and safety related control features will not be interfered by attackers through tampering the sensor data.
\item \emph{Integrity} of the safety-related control operations such as braking and speed control are performed in accordance with the sensed road conditions or from remote control instructions.
\item \emph{Integrity} of the navigation-related control operations such as steering, braking and speed control are performed in accordance with the sensed road conditions or from pre-programming route path.
\item \emph{Confidentiality} of cryptographic keying materials stored inside the AV are ensured so that attackers cannot bypass higher level security mechanisms by siphoning the cryptographic keys.
\item \emph{Integrity} of the kill switch that forces the AV to brake/steer in presence of an unavoidable obstacle, overriding other AV mechanism.
\end{itemize}

\subsubsection{Navigation}
The attacker is capable of mounting an attack through the interface used by a third-party navigation service provider.

\textit{Security-by-Design:} The attacker is able to penetrate the internal network of the AV and thus, get access to unprotected components, which are sensitive to the safety and security of the AV. This can be done by either breach-of-trust (by the navigation service provider), or by the software/hardware interface responsible for receiving navigation inputs, such as GPS receivers.

Note that, in this attack, the RoT assumption is violated since, we do not necessarily assume the navigation service provider as a trusted party anymore. Here, one can ensure robust navigation by designing a \textit{supplementary GPS system}. The objective of this system will be to provide the car a sense of its own position, thus reducing singular dependency on the third party service providers.

To aid this system, the security objective is to ensure - integrity of navigation related control operations. The specific security measures required are essentially same as the ones for collision avoidance. Furthermore, one needs to adopt a spoof identification mechanism/supply an independent GPS system.

\begin{table*}[hbt]
	\centering
	\caption{AV Security-by-Design}
	\label{tab:av-sbd-practice}
	\begin{tabular}{c|c|c|c|c}
		\hline
		\textbf{Safety} 		& \multirow{2}{*}{\textbf{RoT}} 	& \textbf{Security} 	& \multicolumn{2}{c}{\textbf{Security Objective}}\\ \cline{4-5}
		\textbf{Objective}  	&								    & \textbf{Perimeter}	& \textbf{Integrity}  	& \textbf{Confidentiality} \\\hline
		\multirow{2}{*}{Collision Avoidance}	 &	Trusted Third Parties  & \multirow{2}{*}{AV Gateway}	&	Sensor, Navigation-system, &	\multirow{5}{*}{Key Management}					\\
                            &	Key Management			&					&	Kill Switch 		&						\\ \cline{1-4}
		\multirow{2}{*}{Navigation}			&	\multirow{2}{*}{Key Management}			& \multirow{2}{*}{Safety-system}		&	Sensor, Navigation-system, &	\\
                            &                           &                   &   GPS Spoof Detection &	\\ \cline{1-4}
		\multirow{2}{*}{Remote Control}		 &	Trusted Third Parties		& \multirow{2}{*}{AV Gateway}  &	Authentication of & \\
							                 &	Key Management			    &					           &	Remote Control    &	\\\hline
	\end{tabular}
\end{table*}

\subsubsection{Remote Control}
The attacker is able to achieve control of the car network through a remote control component. Consequently, she/he is capable of taking over the cruise/brake control of the AV, which enables her/him to completely disrupt the journey of the AV.

\textit{Security-by-Design:} Here, unlike the navigation system, the attacker looks into the remote control access mechanism and uses those gateways for gaining access to the car network. Once the car is accessible, multiple attacks can be mounted, if the AV does not have proper integrity/authenticity/confidentiality checks in place for an intruder. For a remote keyless entry, it is possible to apply some variant of a replay attack, which records the code (typically from an RF transmitter), and later resends to gain entry to the vehicle.

Note that this attack violates the security perimeter and gain access to the car. To prevent such scenario, the AV can have an onboard \textit{IoT communication monitor}. This device will maintain all sorts of IoT communications that's related to this particular AV. The device can flag a warning when a set of unusual activities are recorded. The specific security measures required are as following.

\begin{itemize}
\item \emph{Integrity} of remote control functions of the AV, e.g., through biometric identification and/or two-factor authentication.
\item \emph{Integrity} of the safety-related control operations such as braking and speed control are performed in accordance with the sensed road conditions or from remote control instructions.
\item \emph{Confidentiality} of cryptographic keying materials stored inside the AV are ensured so that attackers cannot bypass higher level security mechanisms by siphoning the cryptographic keys.
\end{itemize}

\subsection{AV Security Management}
While the security-by-design principles attempt to minimize the security risks in a systematic manner, security management for AV is necessary for continuous assessment of the security challenges during the entire lifetime of an AV. To that effect, we discuss the threat analysis and risk assessment, security testing techniques and legal perspectives of AV security incidents in the following.

\subsubsection*{Threat Analysis and Risk Assessment (TARA)}
Since the standards outlined earlier only help in assigning a quantification under a common assumption of threats, risks and vulnerability -- it is important to assess these factors and undertake a rigorous cybersecurity testing, possibly in an automated manner. It should be noted that the Threat Analysis and Risk Assessment (TARA) is specified as part of SAE J3061, which, however, leaves the determination of the risk level and choice of TARA method to the specific organisation. To that end, there are few prominent methods listed below, for which a detailed review can be found here~\cite{macher_threat_review_2016}.

\begin{itemize}
\item \textit{Attack Tree Analysis:} A standard method is to base on the Attack Trees, as done for the EVITA Threat and Operability (THROP) analysis~\cite{macher_threat_review_2016}. From a functional/feature point of view, the threat is analysed using the attack trees. Further, with the combination of potential threats, the worst-case scenario is identified and the risk is quantified. The severity classification in EVITA is distributed across safety, privacy, financial and operational verticals.

\item \textit{TVRA:} TVRA is a standard approach for Threat, Vulnerability and Risk Assessment for CPS, where the threat is associated with the assets in the system. This was developed for data- and telecommunication applications.

\item \textit{Software Vulnerability Analysis~\cite{krsul_phd_thesis_98}:} This is a technique for assessing the vulnerability of a software code. The idea of software vulnerability stems from the fact that the development and actual environment of a software implementation can differ drastically, more so for AV-like resource-constrained environment and under the influence of a malicious attacker.
\end{itemize}

Apart from the above ones, there are further threat analysis approaches based on the STRIDE framework~\cite{ms_stride}, where, first the threat level is determined, followed by the impact level and eventually, the security level.

\subsubsection*{Security Testing Methods}
\begin{itemize}
\item \textit{Penetration Testing:} Penetration testing is commonly performed as part of a security audit, under the setting of a ``black box" or a ``white box" test subject. In a black box setting, the system details are unknown to the tester, whereas in the white box setting, a powerful adversary is assumed and the system internals are provided to the attacker. The goal of the penetration testing to identify the vulnerabilities and assess the system security.

\item \textit{Red Teaming:} This is a process for detecting network and system vulnerabilities by assuming the role of an attacker, also alternatively termed as ethical hacking.

\item \textit{Fuzz Testing:} In fuzz testing, huge amount of random data is input to the software/system in an attempt to make it crash. The goal is to test coding error and security loopholes.

\item \textit{Network Testing:} In this test, the network resilience is tested by passing large number of packets in short bursts. Further to the stress test, the network configuration and the malicious activities, if any, are tested by performing packet decoding and matching network topology.
\end{itemize}

Given the diverse kinds of TARA approaches and security testing techniques for secure system design, it is necessary to define and adopt an appropriate methodology for AV security management, especially considering both the OT aspects (repairing/servicing) as well as the IT aspects (software update, malicious nodes).

\section{Conclusion and Open Problems}
Security of AV is an important issue and should be considered in a systematic and holistic manner. In this paper, we argued that the security-by-design principle for AV is poorly understood and rarely practiced. We addressed the issue by modeling an AV as a cyber-physical system and studied the AV security objectives by viewing from the perspective of a socio-technical framework. This was done methodically by developing a security-by-design framework for AV from the first principle.

We derived the security objectives and the necessary control measures from the perspective of safety requirements of AV. We argued that one of the key objectives of cybersecurity of AV is to ensure that safety operations are resilience in the face of cyber attacks. Subsequently, the technical challenges and the proposed approaches for AV security were identified and discussed.

Nevertheless, apart from safety assurance, in order for AV to be adopted as a preferred means for transport, the legal and liability issues behind AV remain a major challenge. In essence, technical designs and control measures should be developed to enable law-enforcement agencies and judicial officers to determine the liabilities and the parties at fault in the unfortunate situation of car crashes which could possibly lead to loss of lives. The legal and liabilities issues are important problems that should be addressed as part of the future studies of AV security.

\bibliographystyle{abbrv}
\bibliography{refs,acmart}

\begin{thebibliography}{10}

\bibitem{av_sae_classification}
{Automated Driving: Levels of Driving Automation as per SAE International
  Standard J3016}.
\newblock \url{https://www.sae.org/misc/pdfs/automated_driving.pdf}.

\bibitem{nist_sp800_160}
{National Institute of Standards and Technology Special Publication (SP)
  800-160, Systems Security Engineering: Considerations for a Multidisciplinary
  Approach in the Engineering of Trustworthy Secure Systems, November 2016}.
\newblock \url{https://doi.org/10.6028/NIST.SP.800-160}.

\bibitem{nist_sp800_64}
{National Institute of Standards and Technology Special Publication (SP) 800-64
  Revision 2, Security Considerations in the System Development Life Cycle,
  October 2008}.
\newblock
  \url{http://nvlpubs.nist.gov/nistpubs/Legacy/SP/nistspecialpublication800-64r2.pdf}.

\bibitem{ms_stride}
{The STRIDE Threat Model}, 2002.
\newblock
  \url{https://msdn.microsoft.com/en-us/library/ee823878(v=cs.20).aspx}.

\bibitem{iec_61508}
{Functional Safety and IEC 61508}, 2010.
\newblock \url{http://www.iec.ch/functionalsafety/}.

\bibitem{iso_26262}
{ISO 26262-1:2011}, 2011.
\newblock \url{https://www.iso.org/standard/43464.html}.

\bibitem{iso_27034}
{SO/IEC 27034:2011+ Information technology — Security techniques —
  Application security}, 2011.
\newblock \url{http://www.iso27001security.com/html/27034.html}.

\bibitem{dagstuhl_2014}
{Socio-Technical Security Metrics}, 2014.
\newblock \url{https://www.dagstuhl.de/en/program/calendar/semhp/?semnr=14491}.

\bibitem{dagstuhl_2016}
{Assessing ICT Security Risks in Socio-Technical Systems}, 2016.
\newblock \url{https://www.dagstuhl.de/en/program/calendar/semhp/?semnr=16461}.

\bibitem{intel_asrw_2016}
{Intel Automotive Security Research Workshops}, 2016.
\newblock
  \url{http://www.intel.com/content/dam/www/public/us/en/documents/product-briefs/automotive-security-research-workshops-summary.pdf}.

\bibitem{berkeley_deep_drive}
{Adversarial Deep Learning for Autonomous Driving}, 2017.
\newblock \url{https://deepdrive.berkeley.edu/node/107}.

\bibitem{iot_asokan_16}
T.~Abera, N.~Asokan, L.~Davi, F.~Koushanfar, A.~Paverd, A.~R. Sadeghi, and
  G.~Tsudik.
\newblock Invited: Things, trouble, trust: On building trust in iot systems.
\newblock In {\em 2016 53nd ACM/EDAC/IEEE Design Automation Conference (DAC)},
  pages 1--6, June 2016.

\bibitem{bechkit_key_predistribution}
W.~Bechkit, Y.~Challal, A.~Bouabdallah, and V.~Tarokh.
\newblock A highly scalable key pre-distribution scheme for wireless sensor
  networks.
\newblock {\em IEEE Transactions on Wireless Communications}, 12(2):948--959,
  February 2013.

\bibitem{Burg_Anupam_Lam}
A.~Burg, A.~Chattopadhyay, and K.~Lam.
\newblock Wireless communication and security issues for cyber-physical systems
  and the internet-of-things.
\newblock {\em Proceedings of the IEEE}, \textit{(to appear)}.

\bibitem{butun_intrusion_detection_14}
I.~Butun, S.~D. Morgera, and R.~Sankar.
\newblock A survey of intrusion detection systems in wireless sensor networks.
\newblock {\em IEEE Communications Surveys Tutorials}, 16(1):266--282, 2014.

\bibitem{Checkoway_Automotive_Attack_2011}
S.~Checkoway, D.~McCoy, B.~Kantor, D.~Anderson, H.~Shacham, S.~Savage,
  K.~Koscher, A.~Czeskis, F.~Roesner, and T.~Kohno.
\newblock Comprehensive experimental analyses of automotive attack surfaces.
\newblock In {\em Proceedings of the 20th USENIX Conference on Security},
  SEC'11, pages 6--6, Berkeley, CA, USA, 2011. USENIX Association.

\bibitem{chu_ibe_wsn}
C.-K. Chu, J.~K. Liu, J.~Zhou, F.~Bao, and R.~H. Deng.
\newblock Practical id-based encryption for wireless sensor network.
\newblock In {\em Proceedings of the 5th ACM Symposium on Information, Computer
  and Communications Security}, ASIACCS '10, pages 337--340, New York, NY, USA,
  2010. ACM.

\bibitem{dolev_yao_it_83}
D.~Dolev and A.~Yao.
\newblock On the security of public key protocols.
\newblock {\em IEEE Transactions on Information Theory}, 29(2):198--208, Mar
  1983.

\bibitem{faezipour_intravnetwork_12}
M.~Faezipour, M.~Nourani, A.~Saeed, and S.~Addepalli.
\newblock Progress and challenges in intelligent vehicle area networks.
\newblock {\em Commun. ACM}, 55(2):90--100, Feb. 2012.

\bibitem{garcia_otero_2010}
M.~Garc{\'i}a-Otero, T.~Zahariadis, F.~{\'A}lvarez, H.~C. Leligou,
  A.~Poblaci{\'o}n-Hern{\'a}ndez, P.~Karkazis, and F.~J.
  Casaj{\'u}s-Quir{\'o}s.
\newblock Secure geographic routing in ad hoc and wireless sensor networks.
\newblock {\em EURASIP Journal on Wireless Communications and Networking},
  2010(1):975607, 2010.

\bibitem{airhopper}
M.~Guri, G.~Kedma, A.~Kachlon, and Y.~Elovici.
\newblock Airhopper: Bridging the air-gap between isolated networks and mobile
  phones using radio frequencies.
\newblock In {\em 2014 9th International Conference on Malicious and Unwanted
  Software: The Americas (MALWARE)}, pages 58--67, Oct 2014.

\bibitem{heiber_privacy_2005}
M.~P. Heiber~T.
\newblock {\em Exploring the relationship between context and privacy}, pages
  35--48.
\newblock Springer Berlin Heidelberg, 2005.

\bibitem{karlof_secure_routing}
C.~Karlof and D.~Wagner.
\newblock Secure routing in wireless sensor networks: attacks and
  countermeasures.
\newblock In {\em Proceedings of the First IEEE International Workshop on
  Sensor Network Protocols and Applications, 2003.}, pages 113--127, May 2003.

\bibitem{6853346_Survey}
S.~K. Khaitan and J.~D. McCalley.
\newblock Design techniques and applications of cyberphysical systems: A
  survey.
\newblock {\em IEEE Systems Journal}, 9(2):350--365, June 2015.

\bibitem{krsul_phd_thesis_98}
I.~V. Krsul.
\newblock {\em Software Vulnerability Analysis}.
\newblock PhD thesis, West Lafayette, IN, USA, 1998.
\newblock AAI9900214.

\bibitem{kwokyan_security_paloalto}
K.~Y. Lam.
\newblock {IoT Security: Cybersecurity from IT to OT}.
\newblock In {\em Navigating the Digital Age: The Definitive Cybersecurity
  Guide for directors and officers (Singapore Edition)}. Palo Alto Networks,
  2016.

\bibitem{Lemke_book_2010}
K.~Lemke, C.~Paar, and M.~Wolf.
\newblock {\em Embedded Security in Cars: Securing Current and Future
  Automotive IT Applications}.
\newblock Springer Publishing Company, Incorporated, 1st edition, 2010.

\bibitem{macher_threat_review_2016}
G.~Macher, E.~Armengaud, E.~Brenner, and C.~Kreiner.
\newblock {\em A Review of Threat Analysis and Risk Assessment Methods in the
  Automotive Context}, pages 130--141.
\newblock Springer International Publishing, Cham, 2016.

\bibitem{flexray_attack}
D.~K. Nilsson, U.~E. Larson, F.~Picasso, and E.~Jonsson.
\newblock {\em A First Simulation of Attacks in the Automotive Network
  Communications Protocol FlexRay}, pages 84--91.
\newblock Springer Berlin Heidelberg, Berlin, Heidelberg, 2009.

\bibitem{audi_security}
M.-B. Peters.
\newblock {Have a Safe Trip}, 2017.
\newblock \url{https://audi-encounter.com/en/car-security}.

\bibitem{6096958_Survey}
J.~Shi, J.~Wan, H.~Yan, and H.~Suo.
\newblock A survey of cyber-physical systems.
\newblock In {\em 2011 International Conference on Wireless Communications and
  Signal Processing (WCSP)}, pages 1--6, Nov 2011.

\bibitem{Syverson_onion_routing_2001}
P.~Syverson, G.~Tsudik, M.~Reed, and C.~Landwehr.
\newblock Towards an analysis of onion routing security.
\newblock In {\em International Workshop on Designing Privacy Enhancing
  Technologies: Design Issues in Anonymity and Unobservability}, pages 96--114,
  2001.

\bibitem{szlapczynski_2005}
R.~Szlapczynski.
\newblock A new method of ship routing on raster grids, with turn penalties and
  collision avoidance.
\newblock {\em Journal of Navigation}, 59(1):27–42, 2005.

\bibitem{uav_collision_10}
S.~Temizer, M.~Kochenderfer, L.~Kaelbling, T.~Lozano-Perez, and J.~Kuchar.
\newblock {\em Collision Avoidance for Unmanned Aircraft using Markov Decision
  Processes*}.
\newblock American Institute of Aeronautics and Astronautics, 2017/11/12 2010.

\bibitem{tuohy_intravnetwork_15}
S.~Tuohy, M.~Glavin, C.~Hughes, E.~Jones, M.~Trivedi, and L.~Kilmartin.
\newblock Intra-vehicle networks: A review.
\newblock {\em IEEE Transactions on Intelligent Transportation Systems},
  16(2):534--545, April 2015.

\bibitem{weiser_iot_99}
M.~Weiser.
\newblock The computer for the 21st century.
\newblock {\em SIGMOBILE Mob. Comput. Commun. Rev.}, 3(3):3--11, July 1999.

\bibitem{Wolf04securityin}
M.~Wolf, A.~Weimerskirch, and C.~Paar.
\newblock Security in automotive bus systems.
\newblock In {\em Proceedings of the Workshop on Embedded Security in Cars
  (ESCAR)}, 2004.

\bibitem{zhang_ibe_vehicular}
C.~Zhang, R.~Lu, X.~Lin, P.~H. Ho, and X.~Shen.
\newblock An efficient identity-based batch verification scheme for vehicular
  sensor networks.
\newblock In {\em IEEE INFOCOM 2008 - The 27th Conference on Computer
  Communications}, April 2008.

\end{thebibliography}

\end{document}